%% file: Strings2000.tex
\documentclass[12pt]{article}
\usepackage{amssymb,epsfig}
\newcounter{multieqs}

\newcommand{\bq}{\begin{equation}}
\newcommand{\fq}{\end{equation}}
\newcommand{\bqr}{\begin{eqnarray}}
\newcommand{\fqr}{\end{eqnarray}}
\newcommand{\non}{\nonumber \\}
\newcommand{\noi}{\noindent}

\def\alp{\alpha}   \def\bet{\beta}    
\def\del{\delta}   \def\eps{\epsilon}

 \def\sig{\sigma}

\def\cM{{\cal M}}

\def\pa{\partial}

\newcommand{\tr}{\mbox{tr}}

\def\hlf{\frac{1}{2}}

\def\ZZ{\mathbb{Z}}

\def\bcomment#1{}

\usepackage{graphics}

\begin{document}

\thispagestyle{empty}
\setcounter{page}{0}


\begin{flushright}
\begin{tabular}{l}

CU-TP-{994} \\

\end{tabular}
\end{flushright}

\bigskip

\begin{center}

{\Large \bf Aspects of Collapsing Cycles}\\

\vspace{12mm}

{\large Brian R. Greene${}^{}${}\footnote{Invited talk
presented at Strings 2000, Ann Arbor, Michigan.}

\vskip 1cm

{${}^{}$ {\em Departments of Physics and Mathematics,
Columbia University \\
New York, NY 10027}} \\[5mm]}
\vspace{5mm}

{\bf Abstract}
\end{center}

\noi

Much has been learned about string theory over the last few years by
studying properties of cycles and branes in a given background geometry.
Here we discuss three situations (quantum volume, attractor flows/D-brane stability,
and dynamical topology change) in which cycles in a Calabi-Yau
background evolve and/or degenerate in some manner, yielding various
insights into aspects of quantum geometry.

\bigskip

\vfill

\begin{flushleft}
\today
\end{flushleft}

\newpage
\setcounter{footnote}{0}

\section{Introduction}

Over the last few years we have learned much by pushing string theory
into extreme geometrical realms. Here we would like to discuss three
recent works that continue this type of exploration. In the first, we
will discuss aspects of quantum volume, focusing on some of the unusual
features of volume as probed by wrapped branes \cite{BGYK, BGCL}. In the
second, we will discuss aspects of attractor flows, seeking to find
qualitative differences between the spectra of large and small volume
Calabi-Yau compactifications \cite{FDBGMR}. In the third, we will
discuss collapsing cycles in the context of topology changing
transitions, seeking to find situations in which fairly generic boundary
conditions force the background Calabi-Yau space to undergo topology
change as a function of the eleventh dimension in M-theory
\cite{BGKSGS}.

\section{Quantum Volume}

We will work in the context of type II string theory compactified on a
Calabi-Yau manifold $M$. If the real K\"{a}hler form on $M$ is denoted
by $J$, the classical area and volume of the nontrivial 2, 4, and
6-cycles of $M$ are determined by the usual expression: $\int_{C_r} J^r$
where $r$ is the complex dimensions of the cycle $C_r$. The question
that naturally arises, though, is: How do these expressions change when
quantum effects are taken into account? Now, in any situation in which a
classical concept is extended into the quantum domain, without further
information, some ambiguity creeps in as many quantum expressions have
the same classical limit. To sharpen our choice of what quantum volume
means, therefore, we should introduce some physical guide. In
perturbative string theory, the contributions of world sheet instantons
provide one such guide. Namely, consider a correlation function between
three states corresponding to elements of $H^{1,1}(M)$. As is well
known, the lowest order $\alpha'$ contribution involves the classical
intersection form on $M$, but there are nonperturbative corrections in
$\alpha'$ coming from string world sheets that wrap around holomorphic
curves in $M$. These contributions are sensitive to the volume of such
curves as they are weighted by the exponential of the worldsheet action,
which takes the form ${\rm exp}(2 \pi i \int_{C_2} B + iJ)$.
If one uses mirror symmetry to determine the {\it exact} value of such
correlation functions \cite{BGRP} (by doing a variation of Hodge
structure calculation on the mirror manifold $W$), one can then adjust
the value of $\int_{C_2} B + iJ $ to ensure agreement with the
calculation on $M$, thereby giving one definition of the quantum volume
of $C_2$. As found in \cite{CDGP}, this approach yields
\begin{equation}
 \int_{C_2} B + iJ = {{\int_{\gamma} \Omega} \over  {\int_{\gamma_0} \Omega }}
\end{equation}
where $\Omega$ is the holomorphic 3-form on $W$ and $\gamma, \gamma_0$
are suitably chosen 3-cycles on $W$.

It is worth noting a few features of this definition of the quantum
volume of 2-cycles on $M$. First, we are naturally led to a notion of
complex volumes. Second, in the large radius limit on $M$, if one sets
the $B$-field to zero, we recover the usual classical values for the
volumes of 2-cycles. Third, in the small radius regime, there are
departures from classical expectations. For instance, as in \cite{CDGP,
PABGDM}, on the quintic three-fold (and other one-parameter Calabi-Yau
examples), there is a {\it lower} bound on the quantum volume of
2-cycles---explicitly, on the quintic, 2-cycle volumes satisfy
\begin{equation}
{\rm Vol}_{\rm quintic}{\rm (2-cycles)}  \ge .6881 .
\end{equation}
while classically the only constraint on such volumes is that they are
non-negative.

The fourth point is simply that this approach, since it relies
on worldsheet instantons in perturbative string theory, is only
applicable to 2-cycles. To go beyond this limitation it is natural to
turn to developments in nonperturbative string theory.

In particular, if we are interested in the quantum volume of a p-cycle,
we can wrap a D-p-brane around it in a BPS manner, and then note that
the mass of this state is a physical probe of the volume of the cycle.
Hence, we can declare that the quantum volume of the p-cycle {\it is}
the mass of the state obtained by such a BPS wrapping of a p-brane,
suitably normalized. Of course, to make this definition useful, one
needs to be able to reliably calculate these masses. For 3-cycles this
is immediate: in Type IIB string theory, we wrap a 3-brane on a
supersymmetric representative $\gamma$ of given 3-cycle homology class,
and the resulting mass is
\begin{equation}
{\rm Mass(3-brane)} = { {|\int_{\gamma} \Omega|} \over { |\int_{M} \Omega \wedge
\overline \Omega|^{1/2}} }.
\end{equation}
For even dimensional cycles on $M$, we can write down a similar exact
formula by invoking mirror symmetry. Namely, mirror symmetry maps
$\oplus H^{2p}(M,\ZZ) \rightarrow H^3(W,\ZZ)$ and hence any even cycle
$C$ on $M$ is mapped to some 3-cycle $\gamma$ on $W$. The mass of the
corresponding BPS wrapped brane is then given by the formula above
evaluated for $\gamma$ on $W$. In turn, this gives us the corresponding
quantum volume of $C$ on $M$. Notice that in the case that $C$ is a
2-cycle, the magnitude of the resulting quantum volume agrees with that
of perturbative string theory, up to an overall normalization factor.

We see, therefore, that everything boils down to (a) being able to
calculate the period integrals of $\Omega$ over arbitrary 3-cycles at
arbitrary locations in the moduli space of a given Calabi-Yau manifold
(and its mirror) and (b) explicitly realizing the map $\oplus
H^{2p}(M,\ZZ) \rightarrow H^3(W,\ZZ)$. The subtlety here is that the
calculations in (a) quickly become involved and the map in (b) is not
known in general. In practice, we deal with this in the following way.

First, we are often interested in understanding where in moduli space
the quantum volume of a given cycle vanishes. If this cycle is an
odd-dimensional cycle, then there is no need for the map in (b). If the
cycle is even-dimensional (which we shall henceforth assume), we can get
by with a weaker form of the map in (b). Namely, if we know the map in
(b) up to an overall scale factor, then this is enough to determine
ratios of quantum volumes, as well as the precise locations of the
zeroes of the quantum volumes. So, if we find a basis  $\gamma_j$ of
$H^3(W)$ which are {\it proportional} to integral cycles (with the same,
generally unknown, overall proportionality factor) then by this
reasoning we are free to use these cycles in the range of the map in
(b). Such cycles, it turns out, are much easier to identify than are
integral cycles. As for the map itself, we make use of the general
insights of \cite{CDGP,DM} which relates 3-cycles with $log^j(z)$
monodromy about a large complex structure point $z=0$ on $W$ (assuming
for simplicity a one-dimensional moduli space) to cycles of complex
dimension $j$ on $M$. Now the subtlety here is that this identification
is true up to admixtures of cycles of lower dimension, and it is
generally a challenge to work out these lower dimensional contributions
explicitly. Hence, when we describe cycles of complex
dimension $j$ this implicitly means up to such undetermined admixtures.
Finally, as for (a), the difficulty in carrying out the period
calculations at arbitrary points in the moduli space, we make use of the
classical theory of Meijer functions which naturally encode all the
required analytic continuations of solutions of Picard-Fuchs equations
which these computations require (for details see \cite{BGCL}).

Let us turn to some examples. For simplicity, let's begin with the
quintic hypersurface whose K\"ahler form we parameterize as $J = se$
where $s$ is a real number and $e$ is an integral generator of $H^2({\rm
Quintic})$. As the quintic hypersurface has $h^{1,1} = 1$, we can think
of $s$ as parameterizing a real slice through the one-complex dimensional
K\"ahler moduli space. Classically, the volumes of the single
homologically nontrivially 2, 4, and 6 cycles are just: 2-cycle volume:
$s$, 4-cycle volume: $5 s^2/2 $; 6-cycle volume: $5s^3/6$. In
particular, note that only at $s=0$ does any cycle vanish, and at that
point {\it every} cycle vanishes.

Quantum mechanically the story is different. By carrying out the
procedure as above, we find that the quantum volumes of the 2, 4, and 6
cycles as a function of $s$ are as given in Figure 2 (where in Figure 1,
for comparison, we plot the classical result). Notice that for the
quantum volumes there is only one location where a cycle vanishes, and
it is not at $s=0$. Moreover, the cycle that vanishes is the 6-cycle
\cite{BGYK, JPAS}, not the 2-cycle as one would have naturally
speculated. This is a bit odd, since although the entire
Calabi-Yau has zero quantum volume at this point, the 2 and 4 cycle have
nonzero volumes, as they are bounded below by positive numbers
throughout the moduli space. In \cite{BGCL, CL} this phenomenon has been
found to hold in a range of other one parameter examples, so appears to
be generic.

\vskip 0.2 in

$\begin{array}{cc}
\scalebox{0.3}{\input{quintic_graph_cl.pstex_t}}&
\scalebox{0.3}{\input{quintic_graph_q.pstex_t}}\\
\begin{array}{c}
\mbox{Figure 1. {\footnotesize  Classical 2, 4, 6 cycle volumes}}\\
\mbox{\footnotesize on the Quintic.}\\
\mbox{\footnotesize  }
\end{array}~~~~~~~
&
\begin{array}{c}
\mbox{    }\\
\mbox{    }\\
\mbox{Figure 2. {\footnotesize  Quantum 2, 4, 6 cycle volumes }}\\
\mbox{\footnotesize on the Quintic.}\\
\mbox{\footnotesize  }\\ 
\mbox{\footnotesize }\\
\mbox{}
\end{array}~~~~~~~~~
\end{array}$~~~~~~~~~~~~~~~~~~~~~~~~~~~~~~~~~~~~~~~~~~~~~~~~~~~~~~~~~~~~~~~

As a second example, let's consider two-dimensional orbifolds. In the
case of a $\ZZ_2$ orbifold, it was shown in \cite{PABGDM} --- using the
perturbative string theory approach to quantum volume---  that there is
a point in the moduli space where a 2-cycle collapses to zero quantum
volume (the orbifold point itself). But for the case of $\ZZ_3$
orbifolds, there was a bit of a puzzle in \cite{PABGDM} because
calculations showed that the 2-cycle does not vanish anywhere. Yet by
virtue of there being a singularity in the moduli space of the
associated physical model, we expect that some state has gone to zero
mass at that point. But which state? Well, with our nonperturbative
extension of quantum volume and our experience with the quintic above,
we can answer that question and, moreover, not be too surprised by the
result. As shown in Figure 3, we see that that the 4-cycle in this
example goes to zero volume at a point in the moduli space, even though
the 2-cycle never does.

\vskip 0.8 in 
\hskip 2in\scalebox{0.3}{\input{orb_graph.pstex_t}}

\begin{center} 
Figure 3. {\footnotesize Quantum 2 and 4 cycles volumes for
a $\ZZ_3$ orbifold.}
\end{center}

Finally, in \cite{BGCL} these calculations have  been extended to
one dimensional loci in two-dimensional moduli spaces, with a range of
similar violations of classical geometric intuition making themselves
apparent, and in \cite{CL} the whole series of one parameter examples
has been studied with special attention paid to an interesting manifestation
of $T$-duality in this context.

\section{Attractor Flows}

In the discussion of the last section, we implicitly assumed that in any
given homology class in $H_3(M)$ or $H_3(W)$ there is a supersymmetric
representative. As a matter of fact, however, the existence of
supersymmetric cycles is a difficult and as yet unsettled question. So,
an interesting an important question to ask is (a) under what conditions
are we assured that the relevant supersymmetric cycles exist thereby
implying that the associated BPS state exists, and (b) are there
examples in which the answer to (a) can be used to qualitatively
distinguish large radius and small radius compactifications? That is,
might it be that the existence of a supersymmetric cycle in a given
homology class might have a K\"ahler form dependent answer, existing,
say, when the radius is large but not when it is small?

A full answer to these questions is beyond current understanding, but,
nevertheless, much can be gleaned through a variety of techniques. One
such method makes use of {\it attractor flows}.

Recall that in \cite{FKS}, spherically symmetry black hole solutions to
the equations of ${\cal N} = 2$ supergravity were studied and shown to
obey an attractor mechanism: the value of the fields at the horizon of
the hole are largely insensitive to the boundary values of the fields
at spatial infinity. Explicitly, these equations take the form:
\begin{eqnarray}
\partial_{\tau} U &=& - e^U |Z| \\
\partial_{\tau}z^a &=& -2 e^U g^{a \overline b} \partial_{\overline b} |Z|
\end{eqnarray}
where $\tau = 1/r$, $r$ being the radial direction in space, $U$ arises
from the spacetime  metric in the form $ds^2 = -e^{2U} dt^2 + e^{-2U}dx^idx^i$,
$z^a$ are coordinates on the K\"ahler moduli space, $g^{a \overline b}$ is
the inverse metric on the K\"ahler moduli space in these coordinates, and $Z = Z(\gamma) =
{{ \int_{\gamma} \Omega \over \int_M (\Omega \wedge \overline \Omega)^{\hlf} }}$. 
As one can see, these equations relate flows
in moduli space to the spatial profile of the fields in a black hole
background. As such, they provide an interesting and novel link between moduli
space and spacetime physics.

One expects that these BPS solutions are the supergravity
description of the BPS D-brane states in the full string theory.
In the beautiful work of \cite{GM}, these equations were studied for both their physical
and mathematical content and one observation was this. If we study
an attractor flow solution to these equations which has the property that
$|Z|$ vanishes at a {\it regular} point in the moduli space, we
don't expect the corresponding BPS state to exist. The reason
is that $|Z|$ corresponds to the mass of a BPS brane wrapped on $\gamma$. If
it vanishes along a flow, at that point we expect to
have a new massless degree of freedom, and  this is generally accompanied by
a singularity in the moduli space. If there is no such singularity, we
expect that the corresponding BPS state simply does not exist, i.e. the corresponding
supersymmetric cycle in the homology class $\gamma$ does not exist.
Roughly, we associate $|Z| = 0$ with a collapsing cycle, and the latter
give rise to singularities in the moduli space. If there is no singularity,
 the existence of a supersymmetric cycle in the given class is
thrown into question.

This is a natural conjecture, but in \cite{FD, FDP} it was suggested that there
are circumstances in which the physics is more rich. Namely, there might
be a would-be attractor solution in which there is a zero of $|Z|$ at
a regular point (a ``regular zero") but for which the flow crosses a
curve of marginal stability before that point is encountered. At the
crossing point, the state decays and the attractor flow splits. As for
the physical realization of this state, as shown in \cite{FD, FDP}, if $\gamma$
splits into $\gamma_1 + \gamma_2$, the spacetime configuration can be realized by
a spherical shell carrying the charge associated with $\gamma_2$ which surrounds
the charge of $\gamma_1$ sitting at the origin.

\begin{figure}
\begin{center}
{\epsfxsize=3in \epsfbox{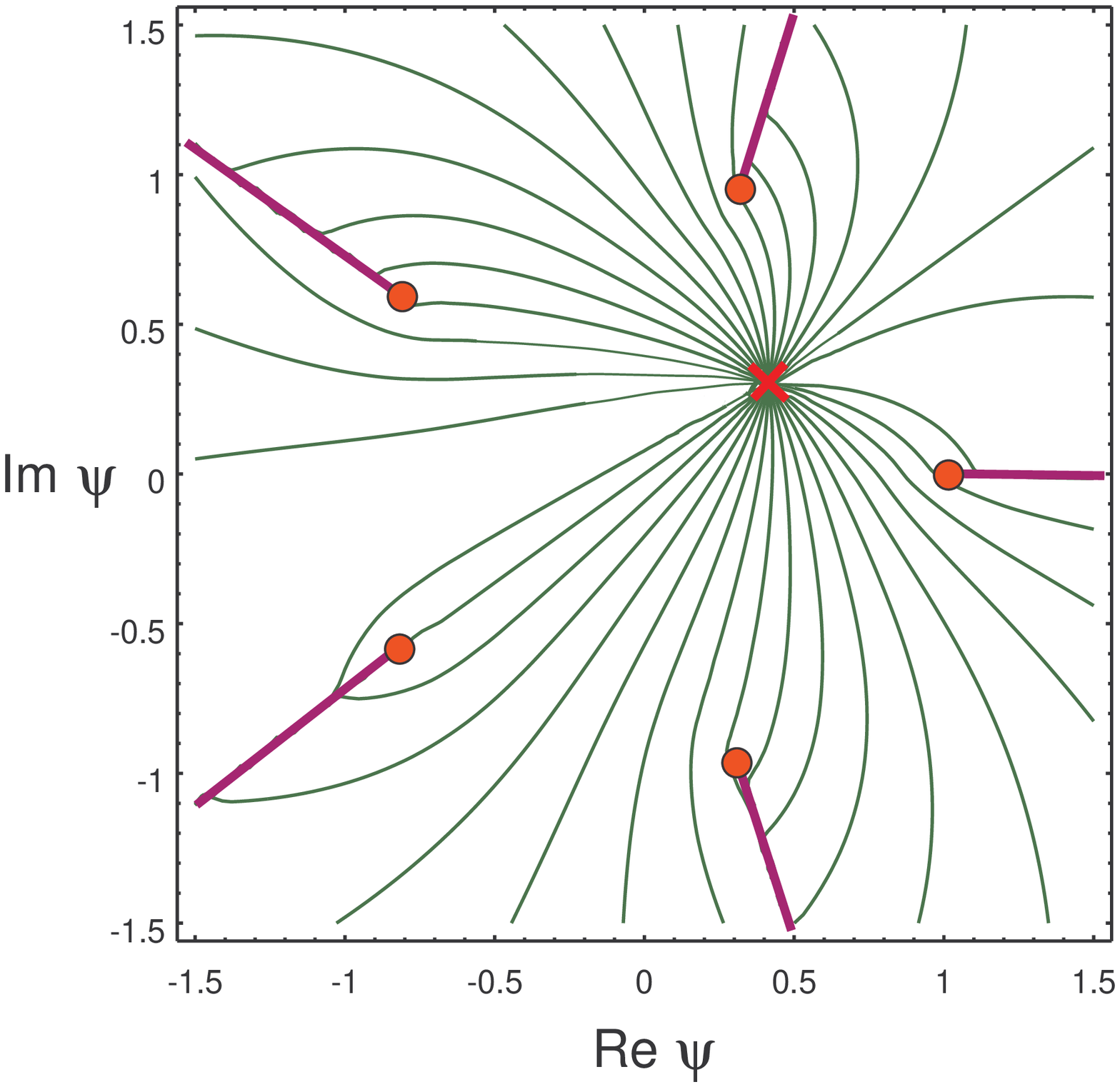}}
\end{center}
\caption{Attractor flows in the quintic moduli space.}
\end{figure}

\begin{figure}
\begin{center}
{\epsfxsize=3in \epsfbox{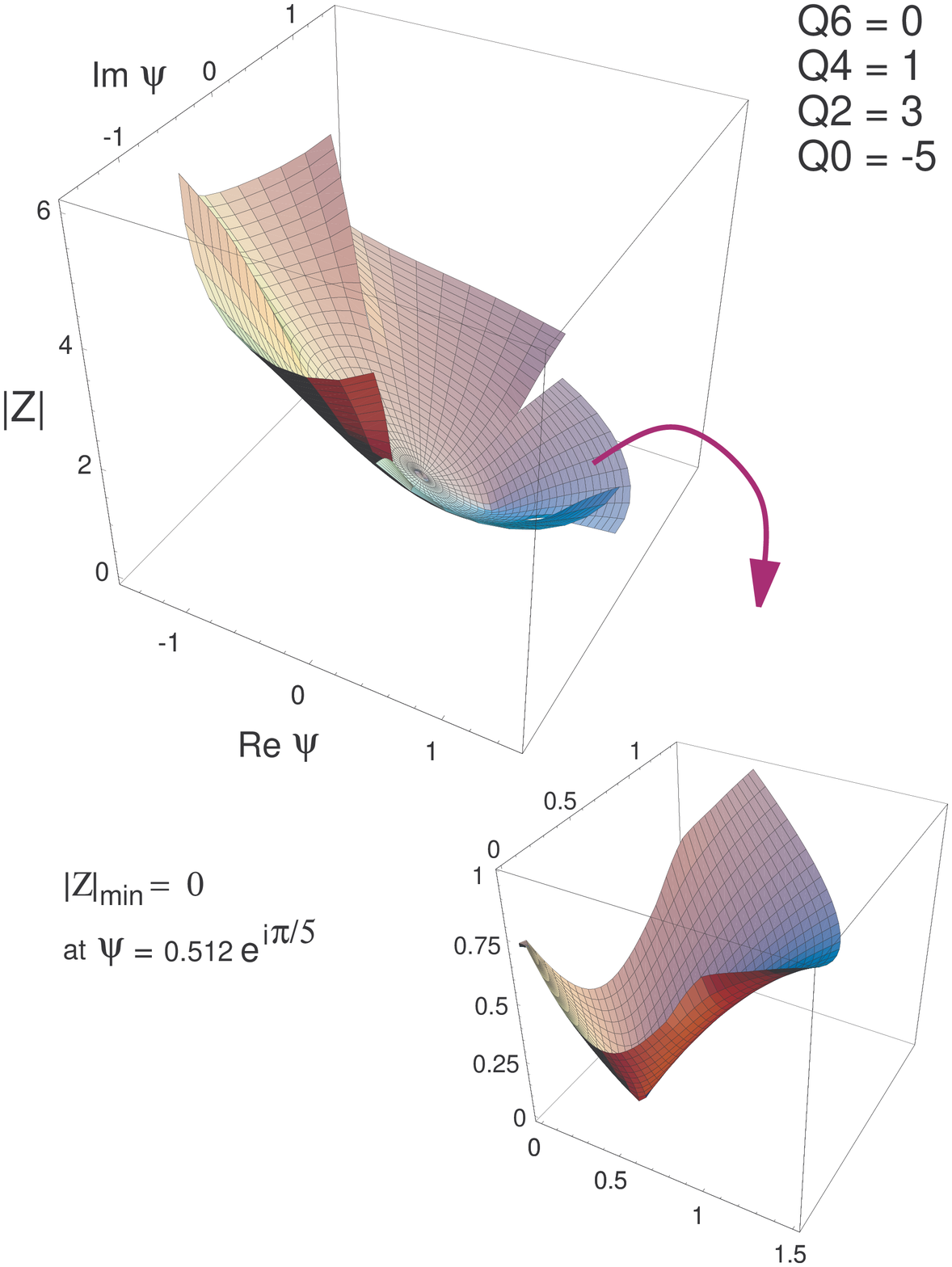}}
\end{center}
\caption{Moduli dependence of the mass of a BPS wrapped brane.}
\end{figure}

An interesting aspect of these solutions, then, is that they provide a means
of identifying candidate states that exist in one region of the moduli space
but not at another. For instance, in \cite{FD, FDP} it was shown  how the  state
identified by \cite{D} --- which exists at the Gepner point on the quintic (which is at
small radius) but apparently decays before getting to large radius --- can be
realized as a splitted attractor flow starting from the Gepner point. Naturally,
then, one wonders about examples in which the reverse would be true: Start at
large radius where classical geometry is a good guide to physics  and identify
states in the string spectrum. Then, go to ever smaller radius where non-classical
geometrical effects become increasingly pronounced. Are there states for   which these
effects cause the state to decay, giving us a qualitative impact of short distance
geometry on  large distance expectations. Indeed in \cite{FDBGMR} such states
have been found,  and in Figures 4, 5, and 6 we give an example.

\begin{figure}
\begin{center}
{\epsfxsize=3in \epsfbox{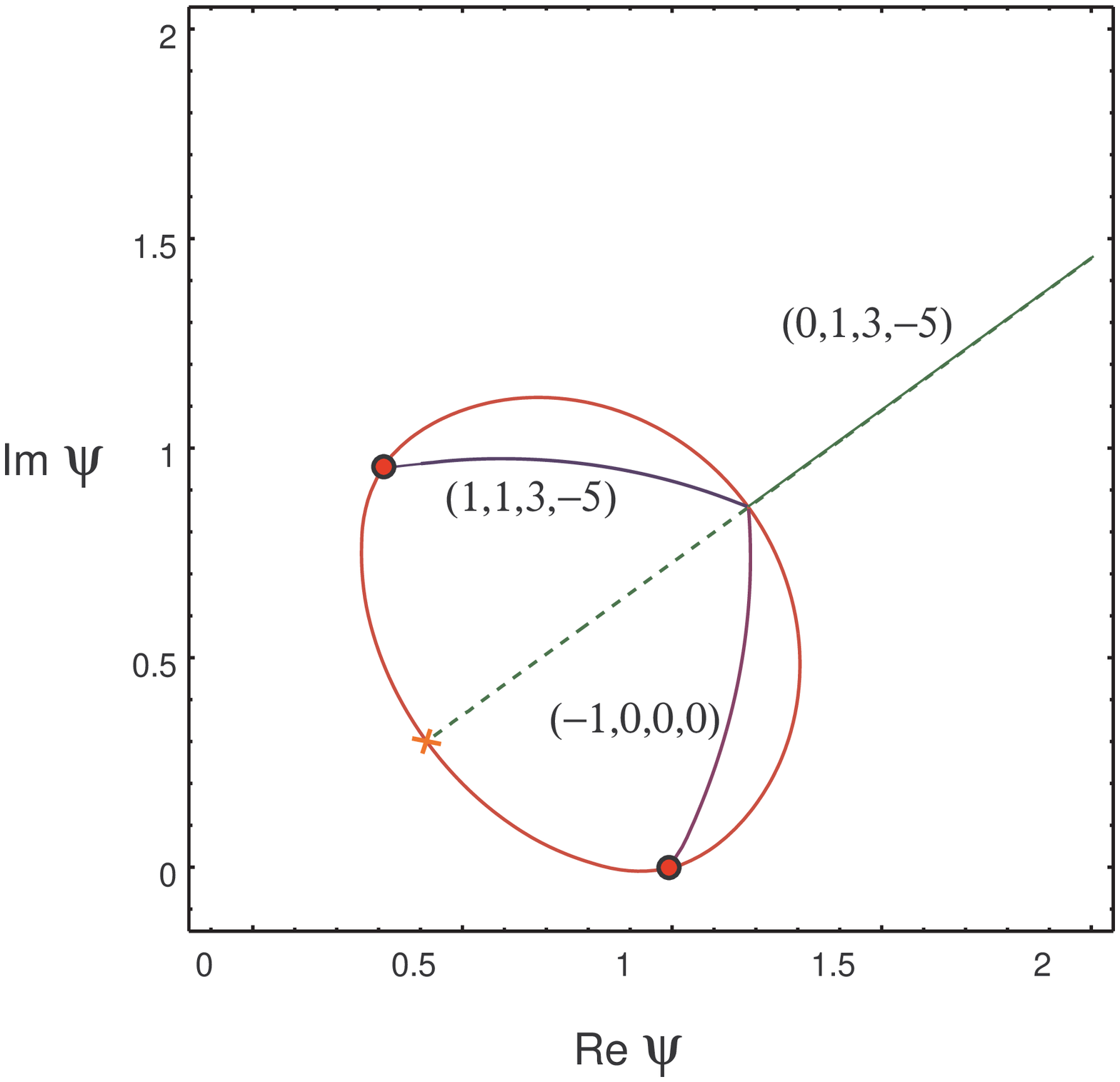}}
\end{center}
\caption{A splitted attractor flow crossing marginal stability line.}
\end{figure}

Namely, in Figure 4 we show a family of attractor flows on the K\"ahler moduli space
of the quintic hypersurface, all associated with the charge vector
$(Q_6,Q_4,Q_2.Q_0) = (0, 1, 3, -5)$. The flows terminate at the attractor point
$\psi = .4146 + .3009i$. As we see in Figure 5, at this point the state
has $|Z| = 0$, but this is a regular point (since the only singular point is
the conifold point, indicated by the red circle(s) in Figure 4). In Figure
6, we show how this  regular  zero is   physically interpreted in
terms of split flow, crossing a curve of marginal stability. This is
but one of many such examples, as discussed in greater detail in
\cite{FDBGMR}, thereby giving a nice qualitative distinction between
long and short distance geometry.

\section{Dynamical Topology Change}

The final arena of collapsing cycles we will discuss has to do with topology
change. Namely, over the years, several works \cite{flop,conifold}
have established definitively
that there are  physically smooth processes in string theory
which result in a change in the topology of spacetime.
In these studies, as well as studies of topology change in M 
theory \cite{Witten:1996qb},
one considers a one parameter family of vacuum
solutions --- a one parameter family of spacetimes --- {that} 
passes from one Calabi-Yau manifold to another which is
topologically distinct. The referenced works succeeded in
showing that there is no obstruction to such topology change, but
no dynamics was ascribed to motion through the family.
In this section, we discuss  \cite{BGKSGS} a variation on this
theme of topology change in which dynamics does
drive the evolution from one topology to another.
Specifically, (a) the topology
change occurs within
a single (not a family of) spacetime background and (b)
for generic choices of initial conditions, the dynamics
({\em i.e.}, the field equations) drive us through
a topology change.

To be concrete, we focus our attention on Calabi-Yau compactifications
of M theory to five dimensions in the presence of $G$ flux
(the four form field strength).
As discussed in \cite{strong,ovrut,gukov,warp}, the effective five
dimensional theory does not admit a flat space vacuum solution.
Rather, the spacetime metric is warped and the solution is
of the domain wall type with one of the five dimensions singled out as 
the transverse direction.
In addition to the effective
five dimensional spacetime metric, the moduli of the Calabi-Yau 
will generically 
vary along the transverse direction.
In \cite{BGKSGS}, we show that there exist Calabi-Yau compactifications 
in which 
the field equations force 
the K\"{a}hler moduli to
pass from one K\"{a}hler cone into an adjacent cone,
while the overall volume of the Calabi-Yau manifold remains large.
This implies that the Calabi-Yau manifold undergoes a flop transition
and continues on 
to a topologically distinct Calabi-Yau manifold
as we move along the transverse dimension.

One may think of this work as being complementary to that
of \cite{blackhole1,blackhole2,lust} in which it was shown that in the
presence of
certain dyonic black holes, a Calabi-Yau with particular
moduli at spatial infinity can be driven by the attractor
 equations through a flop transition on the way to the black
hole's horizon. Here  we briefly describe
vacuum solutions whose structure requires topology change, referring
the reader to \cite{BGKSGS} for details.

Following the pioneering work of \cite{ovrut},
the supersymmetric domain-wall or three-brane solution to the effective
five dimensional field
equations is given by 
\begin{eqnarray}\label{domainwallsoln}
ds_5^2&=&e^{2A}dx_4^2+e^{8A}dy^2 \non
V &=& \left( {1\over 3!} d_{ijk} f^i f^j f^k \right)^2 ~, \nonumber \\
e^{A} &=& V^{1/6} ~, \nonumber \\
b^i &=& V^{-1/6} f^i ~, \non
F^i_{11,\mu\nu\rho\sig} &=&i\sqrt{2}\eps_{\mu\nu\rho\sig}
\pa_{11}V^{-1/2}f^i\
\end{eqnarray}
where the $f^i$'s are defined in terms of one-dimensional harmonic
functions 
\begin{eqnarray}
\label{implicitsoln}
d_{ijk} f^j f^k = H_i ~, \quad \quad H_i &=& \sum_n \alp^{(n)}_i |y-y_n|
+
c_i \\
&=& \sum_{n=0}^{k} 2\alp_i^{(n)} y + k_i~,\quad\quad y_k<y<y_{k+1}
\end{eqnarray}
and the $k_i$ are arbitrary constants of integration. In these expressions,
$V$ is the volume of the Calabi-Yau manifold, the $b^i$ are the K\"ahler moduli,
and $y$ is the coordinate in the ``eleventh" dimension (the coordinate transverse
to the end of the world branes in strongly coupled heterotic string theory).

We see explicitly that the moduli are $y$ dependent, giving rise to
the possibility that in a given example they might pass from one
K\"ahler cone to another as $y$ varies from one wall in spacetime to another.
Indeed this possibility is borne out by studying 
a simple example of a pair of 
Calabi-Yau manifolds connected by a flop transition---namely,
 the well studied
$(h^{1,1},h^{2,1})=(3,243)$ Calabi-Yau manifolds considered in 
\cite{Louis,MorrisonVafa,blackhole1,blackhole2,CFKM}.
Figure 7 shows a topology changing solution to  these equations for this example,
giving rise to the configuration schematically illustrated in Figure 8.
\vskip.2in

\begin{figure}
\vspace*{-.2in}\hspace*{1.75in}
{{\epsfxsize=3.5in \epsfbox{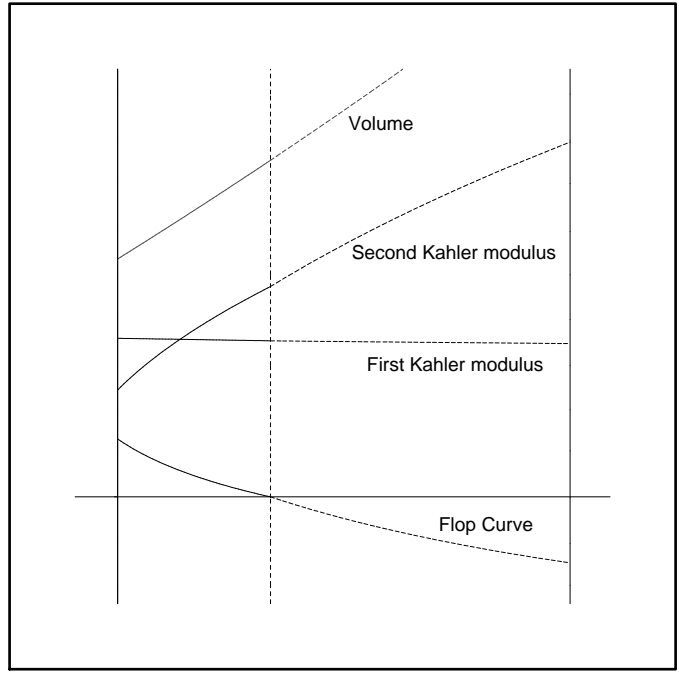}}}
\vspace*{.35in}\caption{Profile of the areas of the two
cycles and the Calabi-Yau volume along $y$. Solid lines belong to the
validity region $y<y_{\ast}$ of the Calabi-Yau $\widetilde{\cM}$. Dashed lines belong to
the
Calabi-Yau $\cM$ and hold for $y>y_{\ast}$.}
\end{figure}

\begin{figure}
\begin{center}
{\epsfxsize=3in \epsfbox{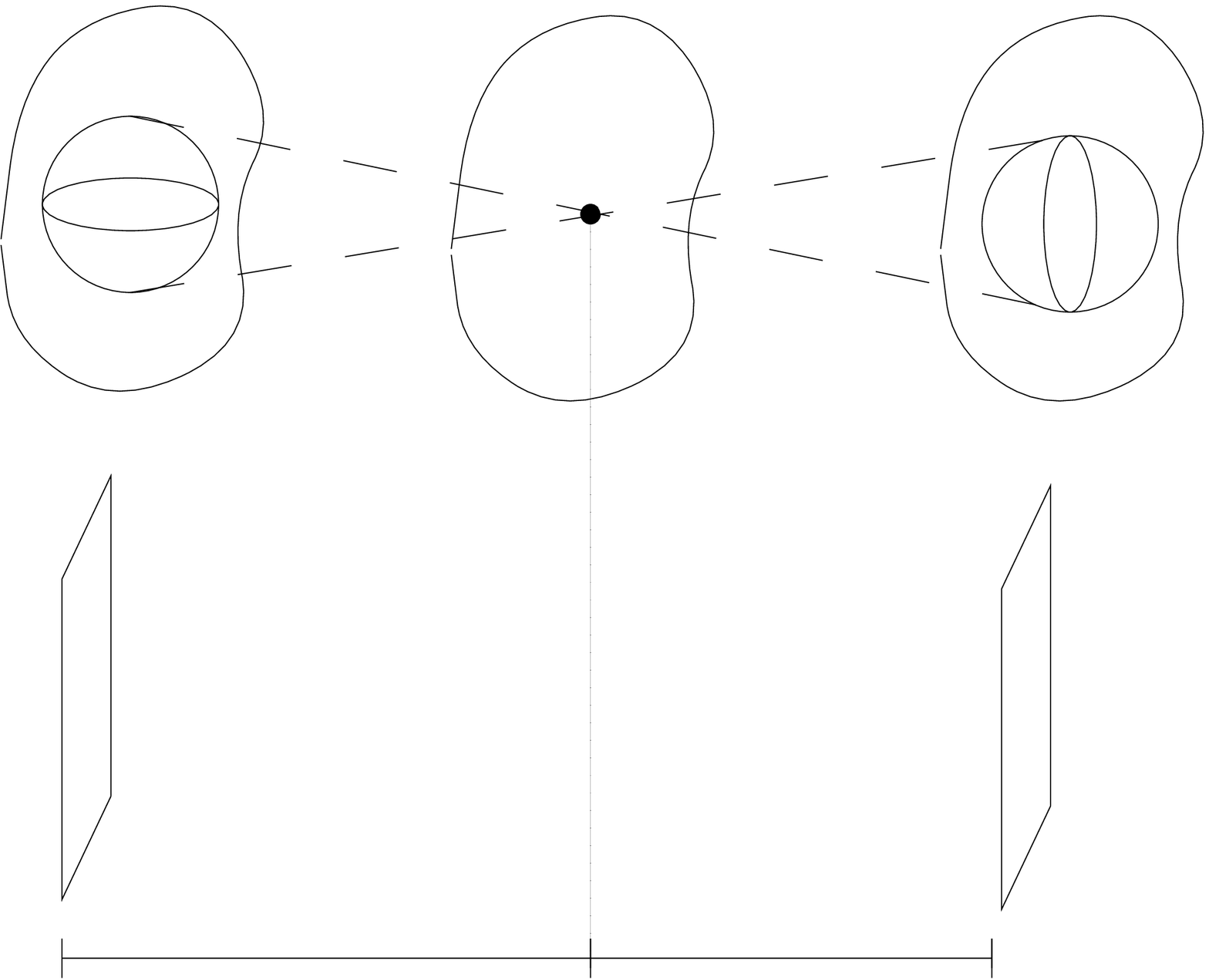}}
\end{center}
\vspace{-2.9in}\hspace{1.7in}{${\widetilde{\cM}}$}\hspace{2.6in}{${\cal
M}$}\vspace{2.6in}
\hspace*{1.7in}{$y=0$}\hspace{0.9in}{$y=y_{\ast}$}\hspace{0.5in}{$y=\pi
R_{11}$}\vspace{.1in}
\caption{Calabi-Yau configuration in strongly coupled heterotic theory
which undergoes a flop.}
\end{figure}

One particularly interesting issue regarding these solutions
arises from examining the Bianchi identity for the G-field.
Namely,  in the eleven-dimensional theory on $S_1/\ZZ_2$,
the Bianchi identity for $G$ is modified by boundary sources
\cite{horava} to the form
\begin{eqnarray}
dG &=& \left( \left[\tr F_{(1)}\wedge F_{(1)}-\hlf \tr R_{(1)}\wedge R_{(1)}
\right]
\del(y) \right.\non 
&& +\left.\left[
 \tr F_{(2)}\wedge F_{(2)}-\hlf
\tr R_{(2)}\wedge R_{(2)}\right] \del(y-\pi R_{11})\right)\wedge dy  ~,
\end{eqnarray}
where we are now taking care to distinguish the curvatures of the
different Calabi-Yau spaces at each end of the interval. In the usual
case, where $\tr R_{(1)}\wedge R_{(1)}=\tr R_{(2)}\wedge R_{(2)}$, we have
the familiar standard 
embedding solution 
\begin{equation}
\tr F_{(1)}\wedge F_{(1)}=\tr R\wedge R~~,~~\tr F_{(2)}\wedge F_{(2)} =0~,
\end{equation}
to the global consistency constraint that
\begin{equation}
\int_{S_1/\ZZ_2 \times D} dG = \int_{S_1/\ZZ_2 \times
D}  dy ~\left(\hlf\tr
R\wedge R ~\del(y) - \hlf\tr R\wedge R ~\del (y-\pi R_{11}) \right) =0.
\end{equation}
But if $\tr R_{(1)}\wedge R_{(1)} \neq \tr R_{(2)}\wedge R_{(2)}$
(cohomologically) then the
mismatch implies that solely embedding either spin connection into the
gauge
group is no longer a solution. 

A natural suggestion, then, is to seek out different holomorphic
stable bundles to  place at $y=0$ and $y=\pi R_{11}$ with different
second Chern classes, so as to find new consistent solutions to the
Bianchi identity. In \cite{BGKSGS}, though, we give evidence that such an approach will
 overlook an essential contribution. Namely, when $\tr R_{(1)}\wedge
R_{(1)} \neq
\tr R_{(2)}\wedge R_{(2)}$ because the Calabi-Yau has flopped somewhere
along $y$,
there is a new contribution to the Bianchi identity associated with the 
collapsed flop curves.

In particular, a theorem of Tian and
Yau \cite{TYau} states that starting from a Calabi-Yau manifold 
$\widetilde{\cM}$ and a collection of holomorphic curves
$\{C^{\bet}\}$ on $\widetilde{\cM}$, the second Chern numbers of
$\tilde{M}$ and its flopped cousin $M$ are related by
\begin{equation}
c_2({\cM})= c_2(\widetilde{\cM})+2\sum_{\bet}
\int_D[C^{\bet}]~,
\label{eq:99}
\end{equation}
with $D$ an arbitrary divisor and 
$[C^{\bet}]\, \eps \, H^4(\widetilde{\cM})$, the Poincare dual of 
$C^{\bet}$. This implies that 
\begin{equation}
-\hlf \tr_{\cM} R_{(2)}\wedge R_{(2)}+\sum_{\bet}\del_{C_{(1)}^{\bet}} = -\hlf
\tr_{\widetilde{\cM}}
 R_{(1)}\wedge R_{(1)}~.
\end{equation}
As suggested in \cite{BGKSGS} this jump should be compensated
by associating a magnetic charge with the collapsed flop curve,
(a charge that might well be interpretable --- in a manner described precisely
in \cite{BGKSGS} --- as that of a 5-brane wrapped on the curve, but
no definitive conclusion on this interpretation was reached).

This work suggests that the study of $G_2$ manifolds with
Calabi-Yau boundaries (a study that has been initiated in
\cite{CHL}) would be a fruitful way of learning more about
these kind of string solutions.

\medskip

\noindent {\bf Acknowledgments}

\smallskip

I would like to thank Calin Lazaroiu, Frederik Denef, Mark Raugas,
Koenraad Schalm, and Gary Shiu with whom I collaborated on
the works presented here. I would also like to thank
 Sergei Gukov,
Shamit Kachru,
Renata Kallosh, Chien-Hao Liu, David Morrison,  Burt Ovrut,
 and
Edward Witten
for useful discussions.
This research  was partially supported by the DOE grant
DE-FG02-92ER40699B.

\end{document}

%% file: quintic_graph_cl.pstex_t
\begin{picture}(0,0)%
\epsfbox{quintic_graph_cl.pstex}%
\end{picture}%
\setlength{\unitlength}{3947sp}%
\begingroup\makeatletter\ifx\SetFigFont\undefined%
\gdef\SetFigFont#1#2#3#4#5{%
  \reset@font\fontsize{#1}{#2pt}%
  \fontfamily{#3}\fontseries{#4}\fontshape{#5}%
  \selectfont}%
\fi\endgroup%
\begin{picture}(10828,7845)(1168,-7723)
\put(8401,-4711){\makebox(0,0)[lb]{\smash{\SetFigFont{20}{24.0}{\rmdefault}{\bfdefault}{\updefault}$s$}}}
\put(8326,-961){\makebox(0,0)[lb]{\smash{\SetFigFont{20}{24.0}{\rmdefault}{\bfdefault}{\updefault}$\frac{5}{6}s^3$}}}
\put(6301,-286){\makebox(0,0)[lb]{\smash{\SetFigFont{20}{24.0}{\rmdefault}{\bfdefault}{\updefault}$\frac{5}{2}s^2$}}}
\end{picture}

%% file: quintic_graph_q.pstex_t
\begin{picture}(0,0)%
\epsfbox{quintic_graph_q.pstex}%
\end{picture}%
\setlength{\unitlength}{3947sp}%
\begingroup\makeatletter\ifx\SetFigFont\undefined%
\gdef\SetFigFont#1#2#3#4#5{%
  \reset@font\fontsize{#1}{#2pt}%
  \fontfamily{#3}\fontseries{#4}\fontshape{#5}%
  \selectfont}%
\fi\endgroup%
\begin{picture}(10828,7845)(1168,-8173)
\put(8326,-6286){\makebox(0,0)[lb]{\smash{\SetFigFont{20}{24.0}{\rmdefault}{\bfdefault}{\updefault}$|t|$}}}
\put(8401,-4711){\makebox(0,0)[lb]{\smash{\SetFigFont{20}{24.0}{\rmdefault}{\bfdefault}{\updefault}$|U_v|$}}}
\put(7201,-1111){\makebox(0,0)[lb]{\smash{\SetFigFont{20}{24.0}{\rmdefault}{\bfdefault}{\updefault}$\frac{2}{25}|U_2|$}}}
\end{picture}

%% file: orb_graph.pstex_t
\begin{picture}(0,0)%
\epsfbox{orb_graph.pstex}%
\end{picture}%
\setlength{\unitlength}{3947sp}%
\begingroup\makeatletter\ifx\SetFigFont\undefined%
\gdef\SetFigFont#1#2#3#4#5{%
  \reset@font\fontsize{#1}{#2pt}%
  \fontfamily{#3}\fontseries{#4}\fontshape{#5}%
  \selectfont}%
\fi\endgroup%
\begin{picture}(10828,7845)(1168,-7723)
\put(7426,-361){\makebox(0,0)[lb]{\smash{\SetFigFont{20}{24.0}{\rmdefault}{\bfdefault}{\updefault}$|U_v|$}}}
\put(8251,-1111){\makebox(0,0)[lb]{\smash{\SetFigFont{20}{24.0}{\rmdefault}{\bfdefault}{\updefault}$|t|$}}}
\end{picture}